\documentclass[5p, sort&compress]{elsarticle} 
\usepackage{graphicx}
\usepackage{amssymb}
\usepackage{epstopdf}
\usepackage[amssymb,thinqspace,thinspace]{SIunits}
\usepackage{amsmath}
\usepackage{booktabs}
\usepackage{rotating}
\usepackage{multirow}
\usepackage{psfrag}
\usepackage{adjustbox}
\usepackage{lineno}
\modulolinenumbers[5]
\usepackage{csquotes}
\usepackage{arydshln}
\usepackage{textgreek}
\usepackage{color}
\usepackage{lipsum,lineno}
\usepackage{float}
\usepackage[normalem]{ulem}
\journal{arXiv}

\begin{document}
\begin{frontmatter}
\title{Tailoring the Deformation Behaviour of a Medium Mn Steel through Isothermal Intercritical Annealing}
\author[1,2]{X. Xu}
\author[1]{T. W. J. Kwok}
\author[3]{P. Gong}
\author[1]{D. Dye\corref{cor1}}\ead{ddye@ic.ac.uk}
\cortext[cor1]{Corresponding author}
\address[1]{Department of Materials, Royal School of Mines, Imperial College London, Prince Consort Road, London, SW7 2BP, United Kingdom}
\address[3]{Department of Materials Science and Engineering, The University of Sheffield, Western Bank, Sheffield, S10 2TN, United Kingdom} 
\address[2]{School of Materials, Sun Yat-Sen University (Shenzhen), and Southern Marine Science and Engineering Guangdong Laboratory (Zhuhai), 519000, China}

\begin{abstract}

A novel concept of varying the strain hardening rate of a medium Mn steel with 8 wt\% Mn by varying the duration of the intercritical anneal after hot rolling was explored. It was found that the stability of the austenite phase showed an inverse square root relationship with intercritical annealing duration and that the maximum strain hardening rate showed a linear relationship with austenite stability. The change in austenite stability was attributed to continuous Mn enrichment with increasing intercritical annealing duration. Twinned martensite was also found to be the most likely product of the martensitic transformation during deformation.

\end{abstract}

\end{frontmatter}


\section{Introduction}
 The latest generation of AHSS, also known as 3\textsuperscript{rd} Generation AHSS (3Gen AHSS), are multiphase steels characterised by simultaneous improvement to strength and ductility through various plasticity enhancing mechanisms such as Transformation Induced Plasticity (TRIP) and Twinning Induced Plasticity (TWIP) \cite{Billur2014c}. More recently, medium Mn steels have shown great promise as candidates for 3Gen AHSS because of their high strength, ductility and strain hardening rate \cite{Hu2017a,Rana2019}. Medium Mn steels are duplex ($\gamma + \alpha$), contain between 4$-$12 wt\% Mn and have been shown to exhibit a combined TWIP$+$TRIP plasticity enhancing mechanism \cite{Lee2015,Ma2017,Rana2019}. 


The key step in the thermomechanical processing of medium Mn steels is the Intercritical Annealing (IA) heat treatment, typically conducted either after hot rolling or cold rolling \cite{Lee2015c,Shao2017a}. The IA heat treatment is conducted between the A\textsubscript{1} and A\textsubscript{3} temperatures, \textit{i.e.} in the $\gamma+\alpha$ temperature regime and the chosen IA temperature determines the equilibrium volume fractions of austenite and ferrite. During IA, solute elements are able to partition between the two phases \cite{Lee2014}. Austenite stabilising elements such as Mn and C diffuse to austenite, and ferrite stabilising elements such as Al and Si diffuse to ferrite \cite{Lee2013c,Hu2017a}. Therefore, the IA temperature also influences the austenite composition and therefore its Stacking Fault Energy (SFE) and stability against martensitic transformation, ultimately governing the resulting alloy's plasticity enhancing mechanism \textit{via} TWIP and/or TRIP mechanisms \cite{Lee2013c,Lee2014}.

One of the challenges facing medium Mn steel is the slow diffusion of Mn to the austenite phase during the IA heat treatment \cite{DeMoor2014,Kamoutsi2015,Nakada2014}. Therefore, the duration of the IA heat treatment is also important so as to achieve the desired Mn content at a given temperature. In the medium Mn steel literature, most researchers approach this challenge in two different ways. The first method is to anneal quickly at a relatively high temperature in a Continious Annealing Line (CAL) \cite{Sohn2014,Lee2011b,Kwok2021} where IA is conducted typically above 700 \degree C and is complete in no longer than 10 min \cite{Granbom2010}. However, this method typically requires a large prior defect density, \textit{i.e.} cold rolled steel, in order to enhance Mn diffusion \cite{Lee2011a,Lee2015e}. In a study by Lee \textit{et al.}, a cold rolling reduction of 66\% (3 $\rightarrow$ 1 mm) was necessary to enable IA to be complete (\textit{i.e.} reach thermodynamic equilibrium) within 180 s. The second method is to anneal slowly at a relatively low temperature in a Batch Annealing Furnace (BAF) for several hours where temperatures are typically below 700 \degree C \cite{Zhang2017a,Lee2017,Field2018}. 

The usage of either a CAL or BAF ultimately aims to drive element partitioning to completion, \textit{i.e.} reaching equilibrium, at a given IA temperature. However, we explore a novel concept of varying the Mn content in the austenite phase through incomplete IA at a constant temperature. If the Mn content of austenite phase can be altered by varying the IA temperature, assuming full partitioning, the same effect can theoretically be achieved by varying the IA duration while the Mn content tends to equilibrium. The range of Mn contents that could be achieved by this method would vary between the bulk Mn and the thermodynamic equilibrium content at the chosen IA temperature. If a sufficiently low IA temperature is chosen, the equilibrium Mn content in the austenite will be relatively high and provide a sufficiently large processing window. Moreover, IA duration, like IA temperature, can significantly influence the recrystallisation and recovery processes during the thermomechanical processing of medium Mn steel and therefore alter the microstructure. This study therefore aims to investigate the effect of IA duration on the tensile properties and microstructural evolution in a medium Mn steel at a constant IA temperature.

\section{Experimental Procedures}

A medium Mn steel with a nominal composition of Fe-8Mn-2.5Al-1.5Si-0.4C-0.02Nb-0.03V in mass percent was arc melted using pure elements to produce an ingot measuring 23 mm $\times$ 23 mm $\times$ 60 mm. The bar was sectioned into four smaller bars measuring approximately 10 mm $\times$ 10 mm $\times$ 60 mm. Each small bar was then quartz encapsulated in low pressure Ar, homogenised at 1200 \degree C for 24 h and then quenched in water. Before hot rolling, each bar was heated to 1000 \degree C for 30 min. The bars were hot rolled in one pass at 1000 \degree C (50\% reduction) and 4 passes at 900 \degree C (25\% reduction per pass) before water quenching immediately after the final pass. The final strip thickness was approximately 1.5 mm. Coupons from the rolled strip were intercritically annealed at 660 \degree C for various durations ranging from 10 min to 24 h before air cooling. Tensile samples were then cut out from each coupon using Electric Discharge Machining (EDM) with gauge dimensions of approximately 1.5 mm $\times$ 1.5 mm $\times$ 19 mm. The tensile samples were cut with the tensile direction parallel to the rolling direction. Tensile testing was conducted at a nominal strain rate of $10^{-3}$ s$^{-1}$. An extensometer was used to measure the strain between 0 and 6-8\%, strain was measured using the crosshead displacement thereafter. \textcolor{black}{One tensile specimen was tested per annealing condition.}

Correlative Electron Backscatter Diffraction (EBSD) and Energy Dispersive Spectroscopy (EDS) were conducted on a Zeiss Sigma FE-SEM equipped with a Bruker EBSD detector and XFlash 6160 EDS detector. Samples for EBSD were first mechanically ground and subsequently polished with an OP-U suspension. Transmission Electron Microscopy (TEM) and Energy Dispersive Spectroscopy (TEM-EDS) were conducted on a JEOL JEM-F200, operated at an accelerating voltage of 200kV. TEM samples were cut from intercritically annealed coupons or post-mortem tensile samples \textit{via} EDM, mechanically ground to a thickness below 60 $\mu$m and electropolished in a Struers Tenupol twin-jet electropolishing unit with a solution containing 5\% perchloric acid, 35\% butyl-alcohol and 60\% methanol at a temperature of $-40$ \degree C.

\begin{table}[t]
	\centering
	\caption{Nominal bulk composition and measured composition of the investigated steel in mass percent. $^\dagger$measured using IGF.}
	\begin{tabular}{lccccccc}
		\toprule
		& Fe    & Mn    & Al    & Si    & C$^\dagger$     & Nb    & V \\
		\midrule
		Nominal & Bal   & 8     & 2.5   & 1.5   & 0.4   & 0.02  & 0.03 \\
		ICP   & Bal   & 8.24  & 2.3   & 1.32  & 0.361 & 0.02  & 0.03 \\
		\bottomrule
	\end{tabular}%
	\label{tab:ICPcomposition}%
\end{table}%

\section{Results}

\begin{figure}[t]
	\centering
	\includegraphics[width=\linewidth]{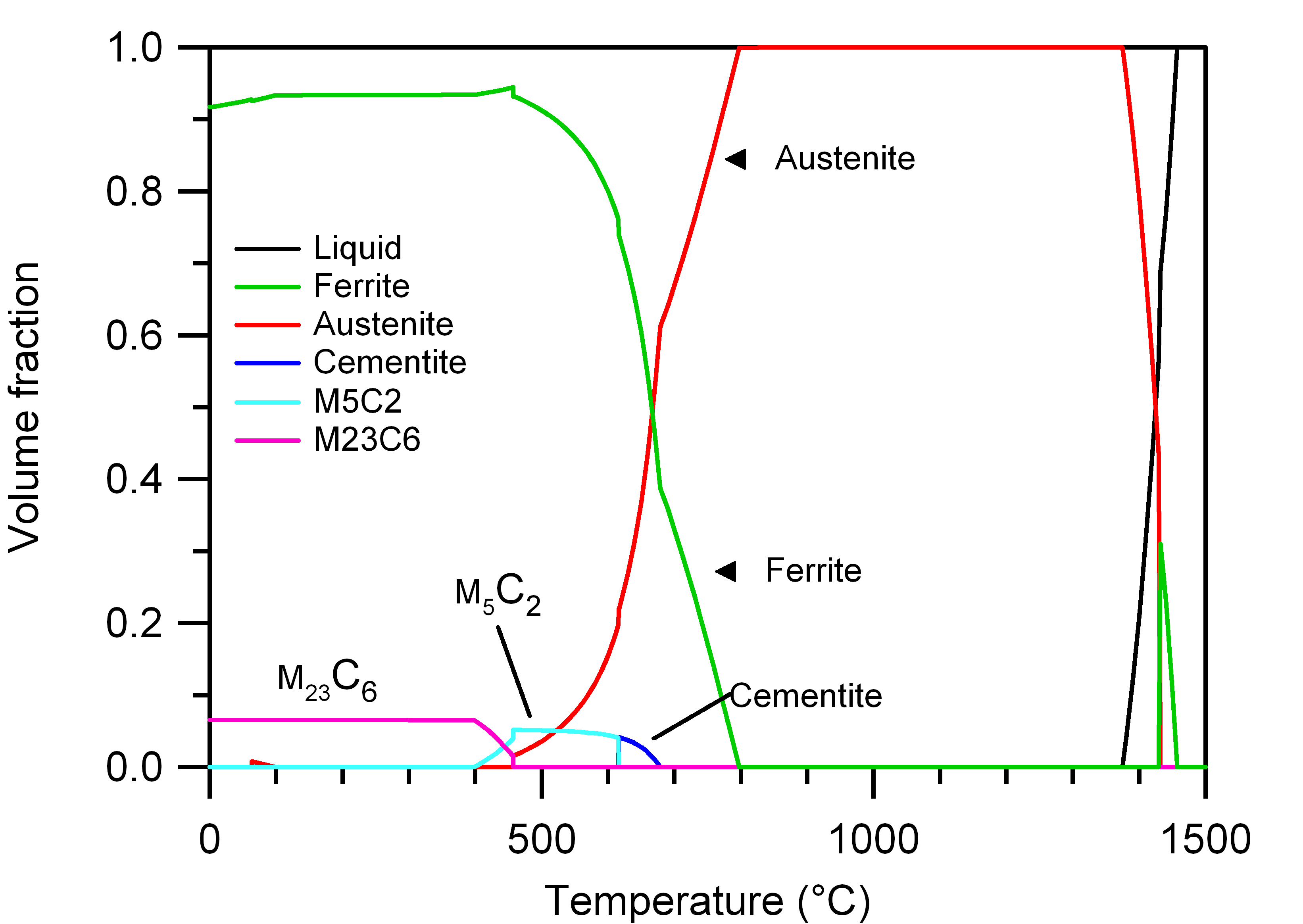}
	\caption{Thermo-Calc property diagram of the measured composition. Some minor phases have been omitted for clarity.}
	\label{fig:fig-nov2-thermocalc}
\end{figure}

The composition of the investigated steel was measured using Inductively Coupled Plasma (ICP) and Inert Gas Fusion (IGF) and is shown in Table \ref{tab:ICPcomposition}. The measured composition was then used to generate a property diagram using the thermodynamic software Thermo-Calc and TCFE7 database in Figure \ref{fig:fig-nov2-thermocalc}. The property diagram shows a wide fully austenitic region (790$-$1375 \degree C) for hot rolling. An intercritical annealing temperature of 660 \degree C was chosen to obtain a microstructure with equal volume fractions of austenite and ferrite and also a small fraction of cementite for carbide precipitation strengthening. It should be noted that in medium Mn steels, the phase fractions and compositions predicted by Thermo-Calc at the IA temperature tend to remain the same at room temperature \cite{Lee2015d,Lee2015c,Kwok2021}.

\begin{figure}[t]
	\centering
	\includegraphics[width=\linewidth]{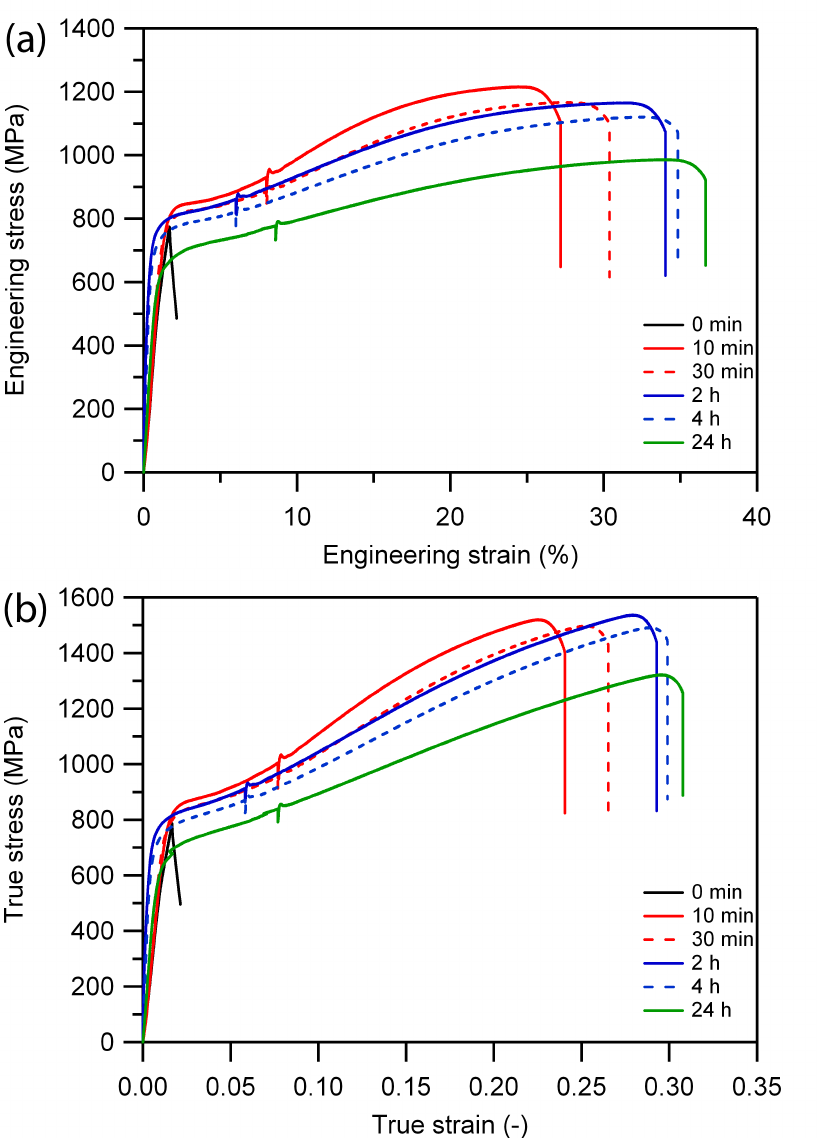}
	\caption{(a) Engineering and (b) true stress strain curves of the investigated steel with varying IA durations. }
	\label{fig:fig-tensiles-engtrue}
\end{figure}

\begin{figure}[h]
	\centering
	\includegraphics[width=\linewidth]{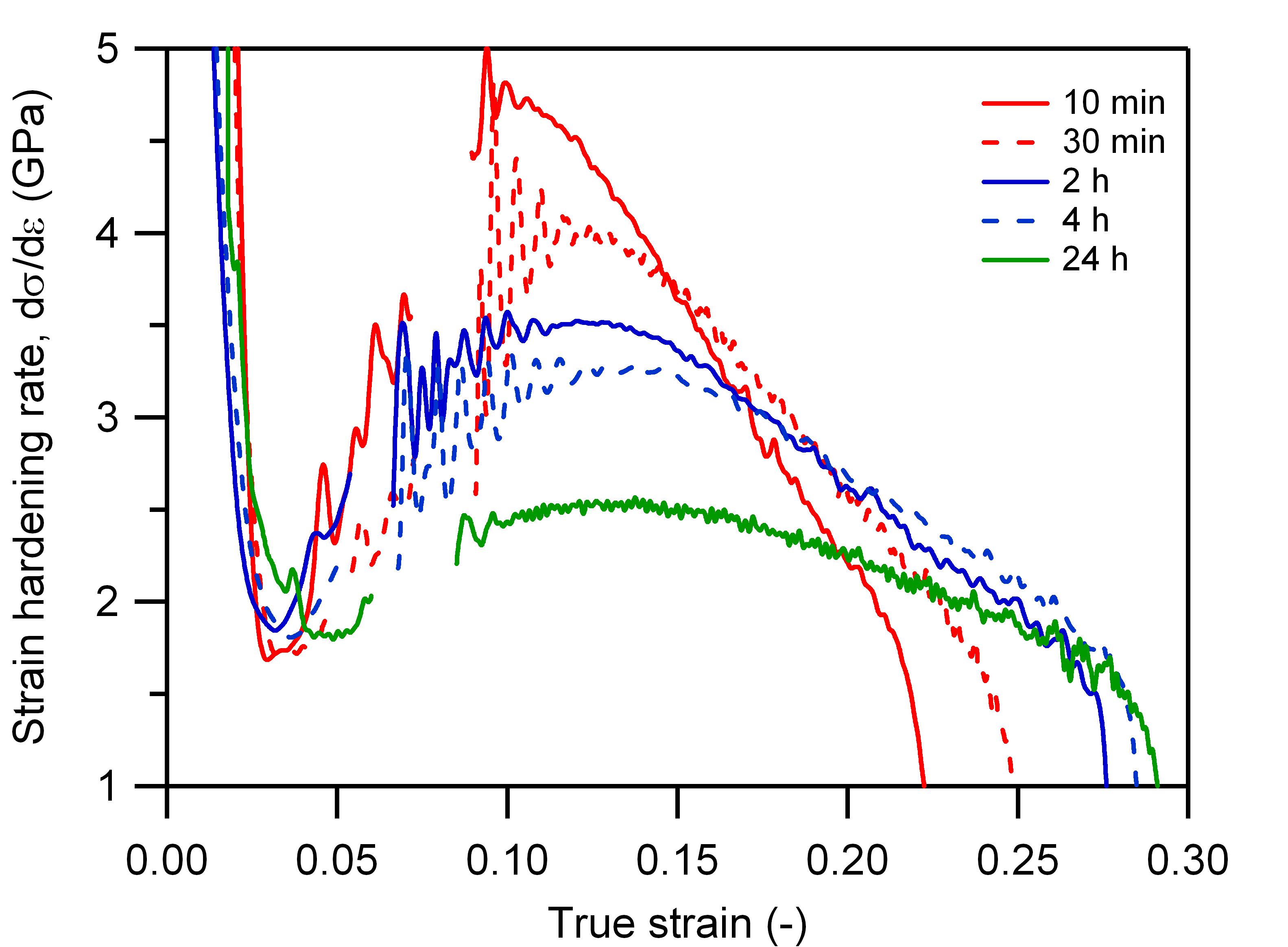}
	\caption{Strain hardening rate of the investigated steel with varying IA duration. A portion of each curve where the extensometer was removed was omitted for the sake of clarity.}
	\label{fig:fig-tensilesshr}
\end{figure}

\subsection{Tensile properties}
The rolled strips were intercritically annealed at 660 \degree C for 0 min, 10 min, 30 min, 2 h, 4 h and 24 h. Henceforth, each sample will be named 660-$X$, where $X$ is the IA duration at 660 \degree C. The tensile curves from each annealing condition are shown in Figure \ref{fig:fig-tensiles-engtrue} and the strain hardening rates of each sample are shown in Figure \ref{fig:fig-tensilesshr}. The tensile properties are summarised in Table \ref{tab:tensileprops}. 

The 660-0 min or the as-quenched sample showed no ductility, fracturing at 724 MPa with only 2\% strain. However, with only 10 mins of intercritical annealing at 660 \degree C, the tensile curve changed drastically. In the 660-10 min sample, the steel exhibited a large improvement in tensile properties with a yield strength ($\sigma_{0.2}$) of 772 MPa, Ultimate Tensile Strength (UTS) of 1216 MPa and ductility of 27\%. When IA duration was increased further up to 24 h, there was a monotonic decrease in yield strength and UTS while there was a monotonic increase in elongation.

From Figure \ref{fig:fig-tensilesshr}, the strain hardening rate, obtained from the first derivative of the true stress-strain curve,  showed a similar shape. After the initial rapid decrease, the strain hardening rate increases to a maximum at approximately 0.1$-$0.15 true strain then decreases up to fracture. However, the maximum strain hardening rate ($\Theta_{max}$) was observed to decrease with increasing IA duration.

\subsection{Evolution of Microstructure}

\begin{figure*}[t]
	\centering
	\includegraphics[width=\linewidth]{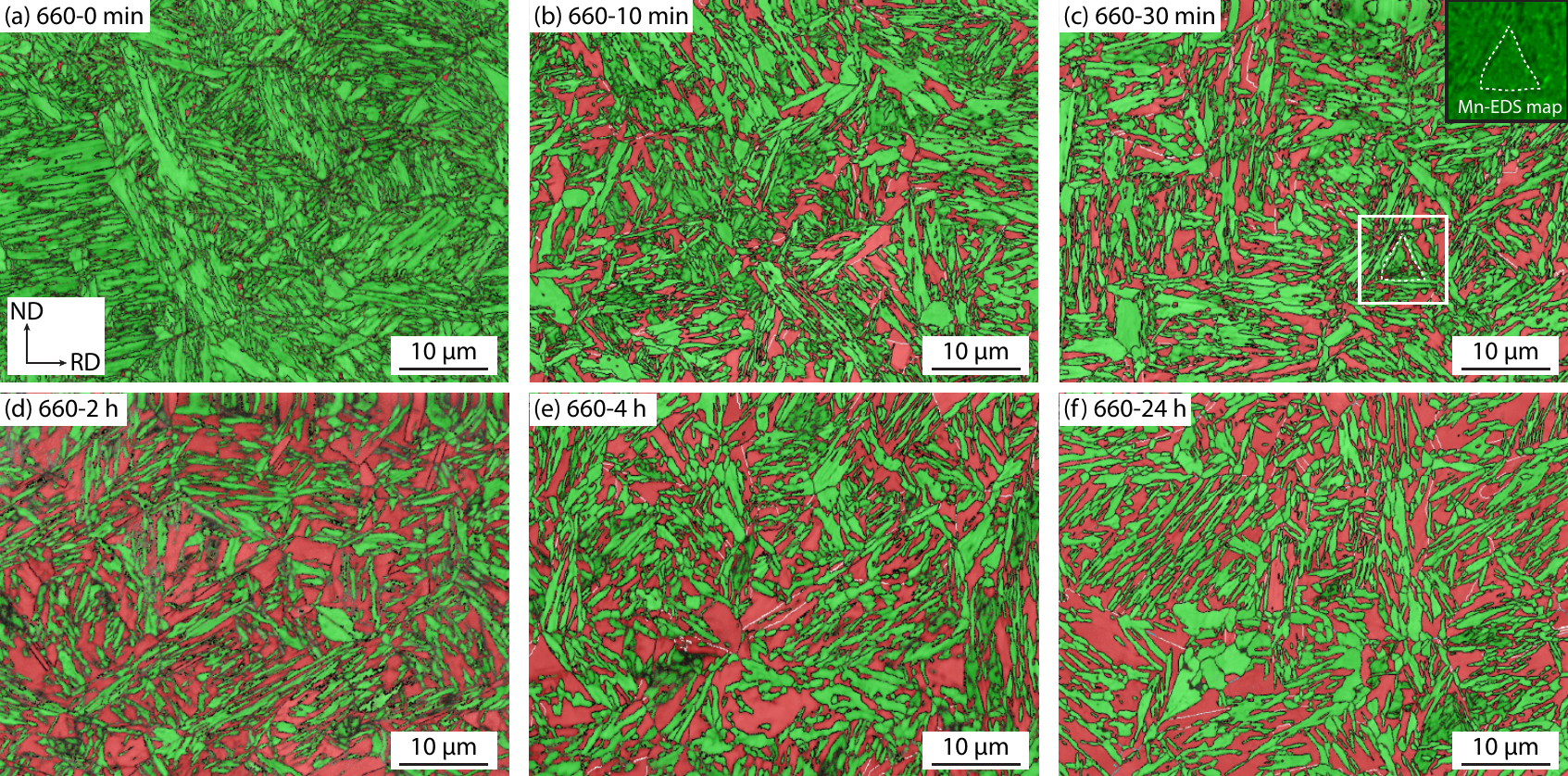}
	\caption{EBSD IQ+PM of the undeformed (a) as-quenched or 660-0 min, (b) 660-10 min, (c) 660-30 min, inset: correlative Mn-EDS map showing that the marked triangular region which indexed as BCC was more enriched in Mn (brighter) compared to the other surrounding BCC grains (darker), (d) 660-2 h, (e) 660-4 h and (f) 660-24 h samples. Red $-$ austenite, green $-$ ferrite/$\alpha'$ martensite. Black lines indicate High Angle Grain Boundaries (HAGBs) and white lines indicate austenite $\Sigma3$ boundaries.  }
	\label{fig:fig-ebsd-undeformed}
\end{figure*}

\begin{table}[htbp]
	\centering
	\caption{Tensile properties of the investigated steel in different annealing conditions.}
	\begin{tabular}{lcccc}
		\toprule
		Sample  & $\sigma_{0.2}$    & UTS   & Elongation & $\Theta_{max}$ \\
		& (MPa) & (MPa) & (\%)  & (GPa) \\
		\midrule
		660-0 min     & 724   & 724   & 2     & NA \\
		660-10 min    & 772   & 1216  & 27    & 4.7 \\
		660-30 min    & 726   & 1167  & 30    & 4 \\
		660-2 h   & 707   & 1165  & 34    & 3.5 \\
		660-4 h   & 676   & 1121  & 35    & 3.2 \\
		660-24 h  & 616   & 986   & 37    & 2.5 \\
		\bottomrule
	\end{tabular}%
	\label{tab:tensileprops}%
\end{table}%

Figure \ref{fig:fig-ebsd-undeformed} shows the change in microstructure with increasing IA duration at 660 \degree C and a summary of the phase fractions obtained through EBSD is shown in Table \ref{tab:phasefracs}. In the 660-0 min sample, the microstructure indexed as predominantly BCC. However, it is likely that the BCC phase was athermal $\alpha'$ martensite rather than $\alpha$ ferrite because of the lack of ductility in the 660-0 min tensile sample. The microstructure was also of the laminate or lamellar-type morphology, which was to be expected in medium Mn steels when quenched directly after hot rolling \cite{Han2017}. 

During IA, the austenite fraction increased rapidly within the first 30 min but slowed down after 2 h before reaching the equilibrium phase fraction as predicted by Thermo-Calc within 24 h. However, when cross referencing the EBSD phase maps with the correlative Mn-EDS maps, it was found that some of the grains which indexed as BCC contained a variation in Mn content. An example is shown in the inset of Figure \ref{fig:fig-ebsd-undeformed}c, where the triangular marked region was enriched in Mn but indexed as BCC, while the surrounding dark (Mn depleted) lamellae also indexed as BCC. It is therefore reasonable to conclude that the triangular region was actually austenite during the IA at 660 \degree C but transformed back to $\alpha'$ martensite during cooling, while the darker laths were intercritical ferrite grains which formed during the IA heat treatment. The intercritical ferrite would be depleted in Mn since Mn partitions out of ferrite into austenite during IA. This suggests that the austenite fraction at 660 \degree C in samples intercritically annealed for less than 24 h might have been larger than that observed at room temperature. In a similar manner, Lee \textit{et al.} \cite{Lee2013c} showed that when the IA temperature was raised too high, \textit{i.e.} forming a high volume fraction of austenite at the IA temperature, some of the austenite would transform back into martensite when cooled to room temperature as the Mn content in the austenite phase became too dilute in order to stabilise it to room temperature. This suggests that the increasing austenite fraction observed in Table \ref{tab:phasefracs} might be due to an increasing Mn content, stabilising more austenite down to room temperature. 


\begin{table}[t]
	\small
	\centering
	\caption{Phase fractions in \% of intercritically annealed samples as predicted by Thermo-Calc and measured with EBSD. N.I. $-$ Non-indexed fraction.}
	\begin{tabular}{lcccccc}
		\toprule
		& $\gamma$ & $\alpha$/$\alpha'$ & $\theta$ & NbC   & VC    & N.I. \\
		\midrule
		660-0 min & 6.7   & 77    &    -   &  -     &   -    & 16.3 \\
		660-10 min & 30    & 65.1  &  -     &   -    &    -   & 4.9 \\
		660-30 min & 33.6  & 62.7  &   -    &   -    &   -    & 3.7 \\
		660-2 h & 41.7  & 35.7  &    -   &    -   &  -     & 22.6 \\
		660-4 h & 42.4  & 56.7  &   -    &    -   &   -    & 0.9 \\
		\smallskip
		660-24 h & 43.8  & 52.6  &  -     &  -     &    -   & 3.6 \\
		Thermo-Calc & 43.7 & 54.2 & 2     & 0.022 & 0.045 & N.A. \\
		\bottomrule
	\end{tabular}%
	\label{tab:phasefracs}%
\end{table}%

\begin{figure*}[t]
	\centering
	\includegraphics[width=\linewidth]{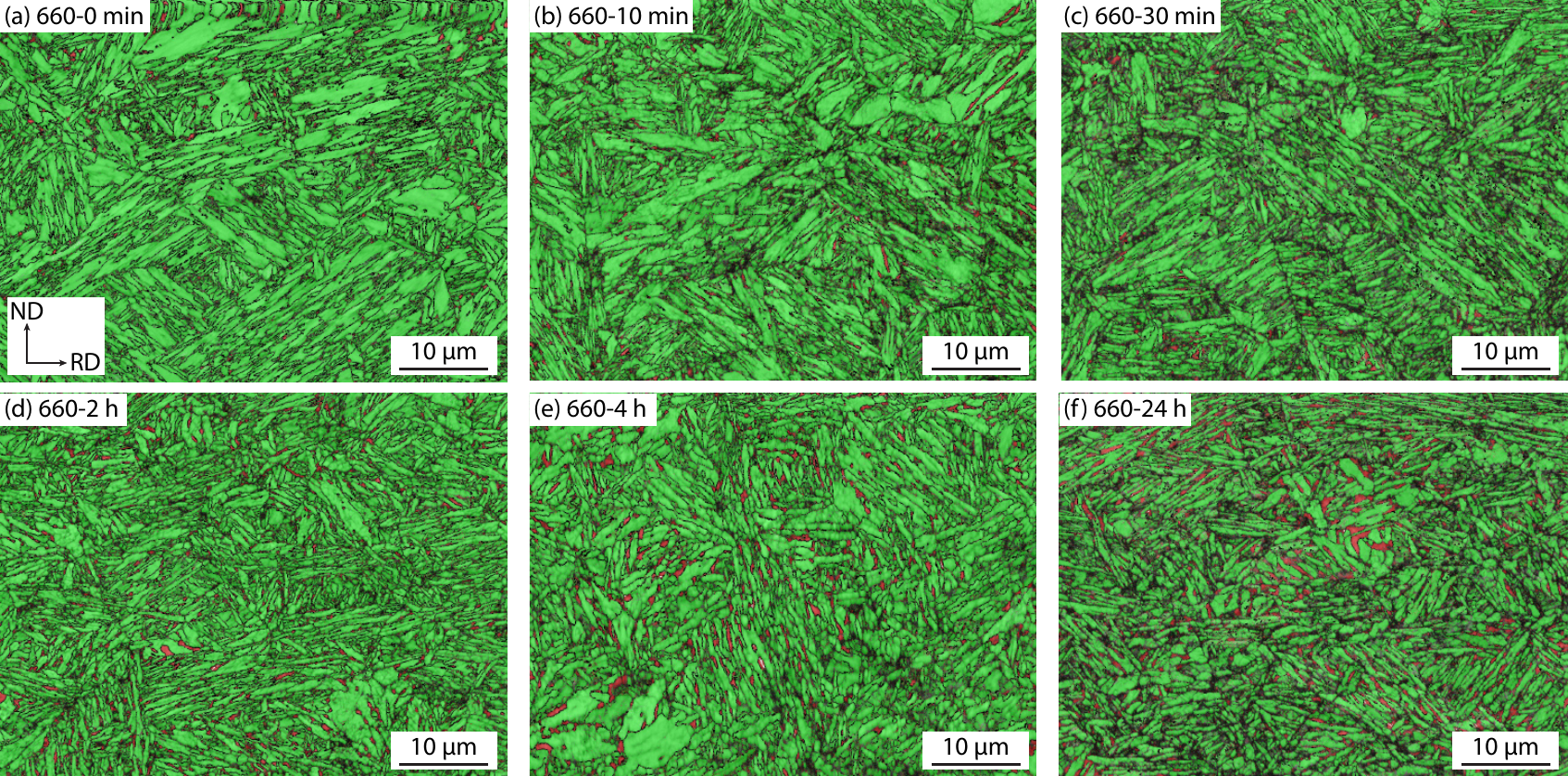}
	\caption{EBSD IQ+PM of the postmortem (a) as-quenched or 660-0 min, (b) 660-10 min, (c) 660-30 min, (d) 660-2 h, (e) 660-4 h and (f) 660-24 h. Red $-$ austenite, green $-$ ferrite/$\alpha'$ martensite. Black lines indicate HAGBs and white lines indicate austenite $\Sigma3$ boundaries.}
	\label{fig:fig-ebsd-deformed}
\end{figure*}

When the tensile samples were deformed to failure, EBSD phase maps were obtained from the \textcolor{black}{gauge of the postmortem samples midway between the fracture edge and shoulder and are} shown in Figure \ref{fig:fig-ebsd-deformed}. The measured phase fractions are shown in Table \ref{tab:phasefracs-postmortem}.  In all the tensile samples, it was observed that the microstructures were almost entirely BCC, \textit{i.e.} ferrite or $\alpha'$ martensite. This shows that a large fraction of the austenite phase had transformed to $\alpha'$ martensite. However, with increasing IA duration, it can be seen that a small but increasing fraction of austenite remained untransformed even after deformation. This suggests that the austenite phase was increasingly stable against transformation with increasing IA duration.  

In order to quantify the austenite stability in medium Mn steels, Steineder \textit{et al.} \cite{Steineder2017a} proposed using the stability constant $k\textsubscript{p}$, after Ludwigson and Berger \cite{Ludwigson1969}. $k\textsubscript{p}$ may be calculated according to Equation \ref{eq:kp_eq1}:

\begin{equation}
\label{eq:kp_eq1}
V_\gamma^{-1}-V_{\gamma0}^{-1} = k_p \: p^{-1} \: \epsilon^p
\end{equation}

where $V_{\gamma0}$ is the austenite volume fraction in the unstrained condition and $V_\gamma$ is the austenite volume fraction at a true strain of $\epsilon$, here taken to be the true strain at fracture. $p$ is a strain exponent corresponding to the autocatalytic effect and can be taken to be equal to 1 as proposed by Matsumura \textit{et al.} \cite{Matsumura1987}. Equation \ref{eq:kp_eq1} can therefore be rewritten as Equation \ref{eq:kp_eq2} where a high $k\textsubscript{p}$ value indicates lower austenite stability and \textit{vice versa}.

\begin{equation}
\label{eq:kp_eq2}
k_p = \frac{V_\gamma^{-1}-V_{\gamma0}^{-1}}{\epsilon}
\end{equation}

\begin{figure}[ht]
	\centering
	\includegraphics[width=\linewidth]{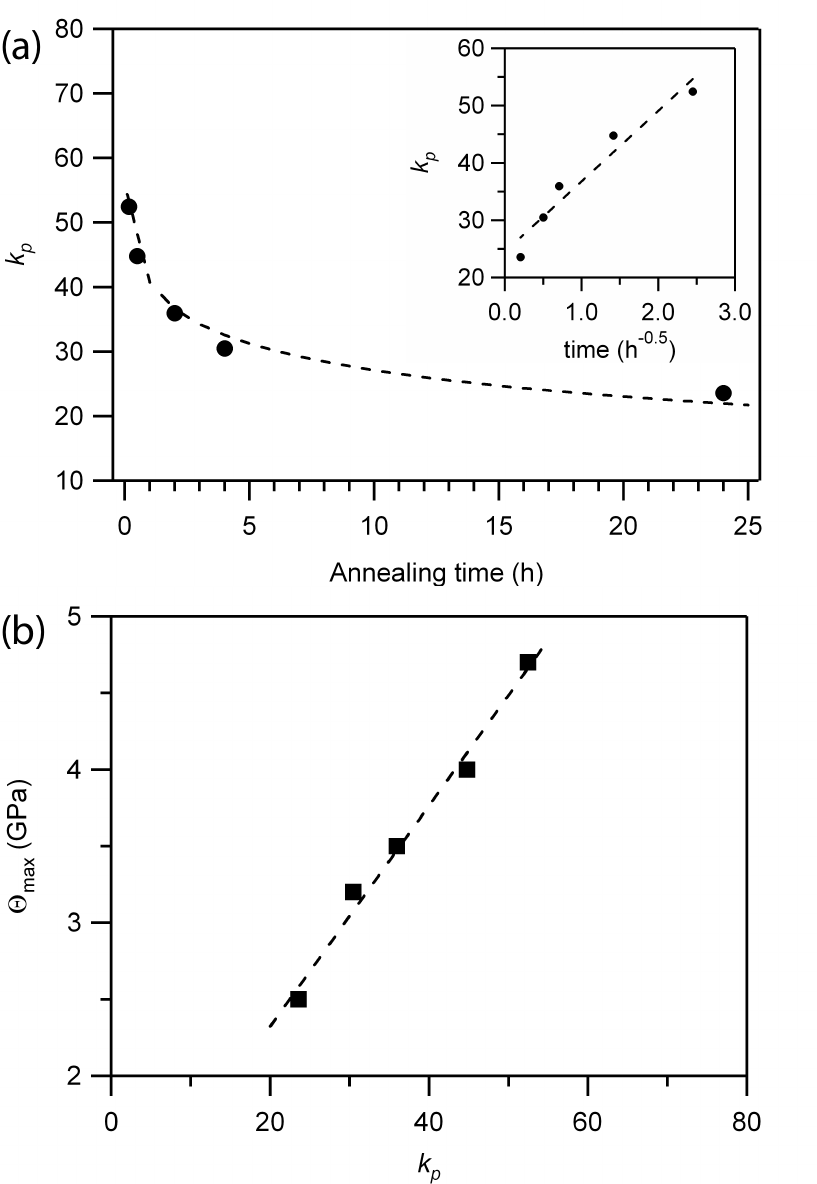}
	\caption{(a) Relationship between $k_p$ and annealing time. Inset: good linear relationship between $k_p$ and the inverse square root of time. (b) A linear relationship is obtained between $k_p$ and $\Theta_{max}$, \textit{i.e.} maximum strain hardening rate.}
	\label{fig:fig-kp-annealing-time-and-shr}
\end{figure}

Based on the austenite phase fractions obtained in Figure \ref{fig:fig-ebsd-undeformed} and \ref{fig:fig-ebsd-deformed}, the change in $k_p$ with annealing time was calculated and shown in Table \ref{tab:phasefracs-postmortem} and Figure \ref{fig:fig-kp-annealing-time-and-shr}a. It can be seen that the $k_p$ value decreased rapidly with time up to approximately 4 h and slowed significantly up to 24 h. However, when $k_p$ was plotted as a function of the inverse square root of IA duration, a good linear fit was obtained as shown in Equation \ref{eq:annealing-time-kp}, where $t$ is in hours.

\begin{equation}
	\label{eq:annealing-time-kp}
k_p =  \frac{12.3}{\sqrt{t}} + 24.5
\end{equation}

In TRIP steels, it is known that the austenite stability strongly affects the kinetics of martensite formation during tensile deformation \cite{Soleimani2020,Chiang2015} and in Figure \ref{fig:fig-kp-annealing-time-and-shr}b, a very good linear relationship was found between the $k_p$ value and $\Theta_{max}$ according to Equation \ref{eq:kp-SHR}: 

\begin{equation}
	\label{eq:kp-SHR}
\Theta_{max} \, (GPa)= 0.07 k_p +0.88
\end{equation}

This strongly suggests that the maximum strain hardening rate in medium Mn steels is highly dependent on the austenite stability even though the initial austenite volume fractions were not the same.

\begin{figure*}[t]
	\centering
	\includegraphics[width=0.8\linewidth]{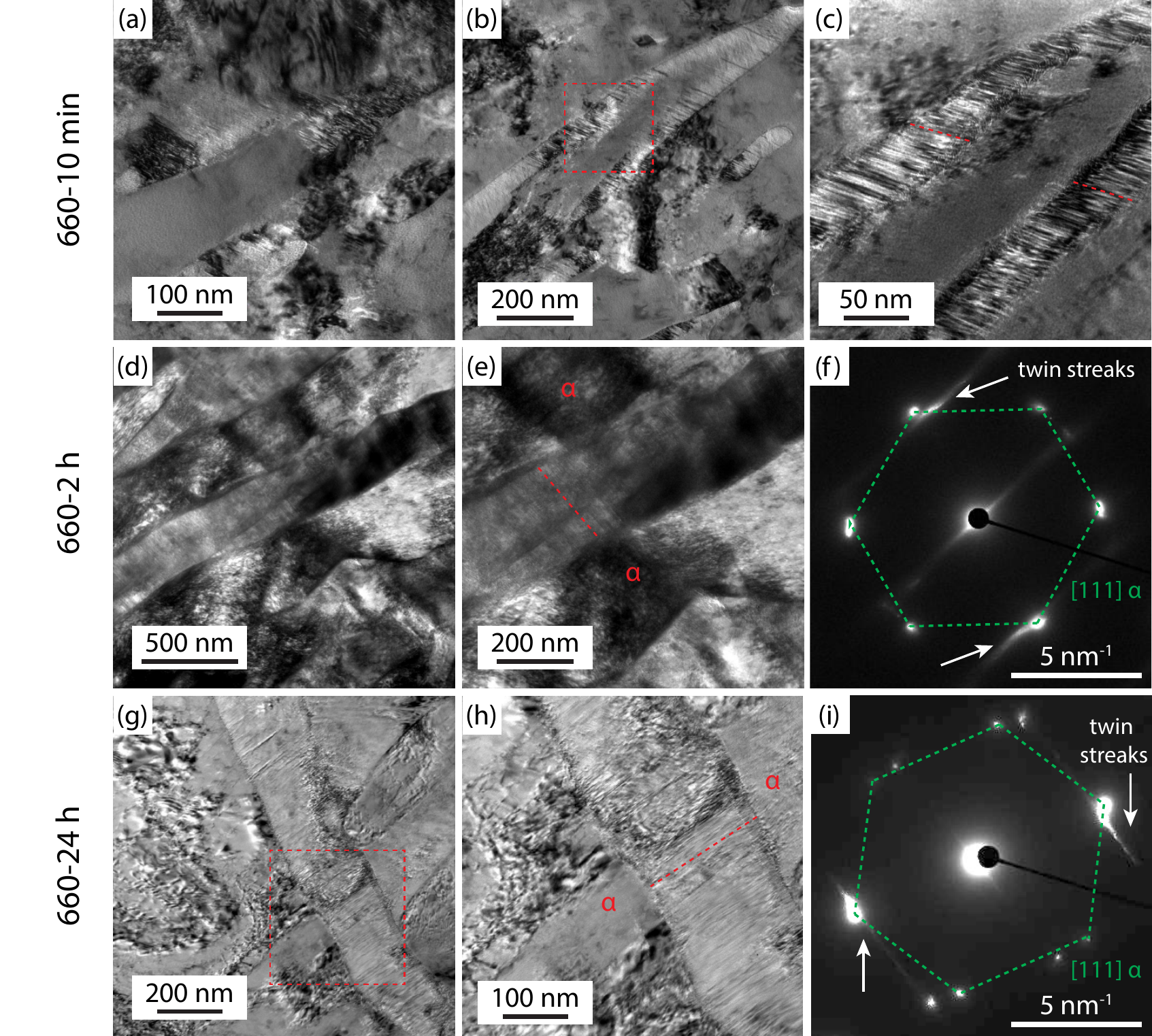}
	\caption{Brightfield TEM micrographs of postmortem samples. (a) Overview of 660-10min, (b) lath structure containing twin-like structures, (c) magnified view of the red square in (b) showing the twin-like structures in greater detail (red dotted lines). (d) Overview of 660-2 h, (e) higher magnification showing very fine twin-like structures along the red dotted line and (f) diffraction pattern obtained from (e) showing twinning streaks. (g) Overview of 660-24 h sample, (h) magnified view of the red square in (g) also showing twin-like structures, (f) diffraction pattern obtained from (e) also showing twinning streaks. Beam direction was parallel to $[110]_\alpha$.}
	\label{fig:fig-tem}
\end{figure*}

In Figure \ref{fig:fig-tem}, the post-mortem samples of 660-10 min, 660-2 h and 660-24 h, were further examined in a TEM. In the 660-10 min sample, very thin twin-like structures were observed inside a lamellar grain. In the 660-2 h and 660-24 h, similar twin-like structures were also found inside lamellar grains. Twinning streaks were observed in the diffraction patterns in Figure \ref{fig:fig-tem}f and i. Therefore, it is likely that the twin-like structures were twinned martensite. Similar twinned martensite were also observed in metastable austenitic stainless steels \cite{Maxwell1974,Krauss1971}.


\begin{table}[t]
	\centering
	\caption{Phase fractions in \% of postmortem samples as measured by EBSD and calculated $k_p$ value.}
	\begin{tabular}{lcccc}
		\toprule
		& $\gamma$ & $\alpha$ & N.I.  & $k_p$ \\
		\midrule
		660-0 min & 4.9   & 84.7  & 10.4  & 276.9 \\
		660-10 min & 6.3   & 74.8  & 18.9  & 52.5 \\
		660-30 min & 6.79  & 62.5  & 30.7  & 44.8 \\
		660-2 h & 7.74  & 75.8  & 16.5  & 36.0 \\
		660-4 h & 8.69  & 76.2  & 15.1  & 30.5 \\
		660-24 h & 10.3  & 53.9  & 35.8  & 23.6 \\
		\bottomrule
	\end{tabular}%
	\label{tab:phasefracs-postmortem}%
\end{table}%



\subsection{Composition}

The types of plasticity enhancing mechanism in medium Mn steels are usually attributed to both the SFE and stability of the austenite phase, both of which are composition and grain size dependent \cite{Lee2014, Saeed-Akbari2009, Angel1954}. In Table \ref{tab:austenite compositions}, the compositions of four samples were measured using TEM-EDS and compared to the equilibrium composition obtained by Thermo-Calc. \textcolor{black}{Based on previous work by Kwok \textit{et al.} \cite{Kwok2022a}, the Mn distribution in the vacuum arc melted ingot was assumed to be homogeneous after the homogenisation heat treatment of 1250 \degree C for 24 h and therefore not susceptible to Mn banding \cite{DeCooman2004}. A minimum of six point scans were obtained for the austenite and ferrite phases each.} The composition of the austenite in the 660-0 min sample was assumed to be the bulk composition, since no partitioning could have occured. C content was determined using the lever rule based on the assumptions that ferrite has negligible C solubility \cite{Lee2014,Sun2018} and C in the form of cementite (0.153 wt\% C) would have fully precipitated within 10 min with the volume fraction as predicted by Thermo-Calc in Table \ref{tab:phasefracs}. Cementite is known to precipitate rapidly even after 1 min of IA \cite{Luo2011} and while Si is known to retard cementite formation, the large C supersaturation is expected to overcome any kinetic effects imposed by Si \cite{Kozeschnik2008}. Grain size was calculated using the Equivalent Circle Diameter (ECD) obtained from the EBSD software ESPRIT. The SFE was calculated according to the method proposed by Sun \textit{et al.} \cite{Sun2018} and Md\textsubscript{30}, a measure of austenite stability against strain-induced transformation, was calculated according to Equation \ref{eq:Md equation} \cite{Angel1954,Nohara1977, Sun2018}:

\begin{equation}
\label{eq:Md equation}
\begin{split}
\small
Md_{30} (\degree C)= \,& 551 - 462 C  - 8.1 Mn  - 9.2 Si \\
& - 1.42(-3.29-6.64\log_{10}d_\gamma-8)
\end{split}
\end{equation}

\noindent
where the compositions are given in mass percent and $d_\gamma$ is the austenite grain diameter. Md\textsubscript{30} is defined as the temperature at which half of the austenite phase transforms to martensite at a true strain of 0.3 \cite{Angel1954}. This parameter was primarily developed to measure the stability against strain induced martensitic transformation in stainless steels \cite{Angel1954,Nohara1977} but has been increasingly used in medium Mn steels too \cite{Sun2018,Hu2019,Kwok2019,Kwok2022b}. A low Md\textsubscript{30} indicates a high austenite stability and \textit{vice versa}. 




From Table \ref{tab:austenite compositions}, it could be seen that the Mn content of the austenite phase in the 660-10 min sample had largely the same Mn content as the bulk composition but rose continuously from the 660-30 min sample to the 660-24 h sample. The rise in Mn only after 10-30 min of intercritical annealing and the rate of growth of austenite fraction in Table \ref{tab:phasefracs} are in agreement with the findings by Kamoutsi \textit{et al.} \cite{Kamoutsi2015} and Farahani \textit{et al.} \cite{Farahani2015}, which showed that austenite initially forms rapidly from martensite under Negligible Partitioning Local Equilibrium (NPLE) mode where growth is controlled by C diffusion. But the rate of austenite growth slows down as it switches to Partitioning with Local Equilibrium (PLE) mode where austenite growth is controlled by Mn diffusion in both austenite and ferrite. Thermo-Calc also predicted that the equilibrium austenite Mn content at 660 \degree C would be 12.5 wt\%, as compared to 11.9 wt\% in the 660-24 h sample. This suggests that the Mn content may not have reached the equilibrium composition, even after 24 h of intercritical annealing. Al, which was expected to partition to ferrite, varied significantly between the measured samples, likely due to some degree of interference with the TEM sample holder. Reasonable conclusions therefore could not be drawn from the Al content which would affect SFE calculation. Nb was undetected in the 660-30 min, 2 h and 24 h samples, suggesting that Nb had completely precipitated in the form of carbides. However, V largely remained in solution. 


Regarding austenite stability, Samek \textit{et al.} \cite{Samek2006} showed that in multiphase TRIP steels, a decreasing $k_p$ should also correlate with a decreasing Md\textsubscript{30}. However, this was not the case in Table \ref{tab:austenite compositions} which showed an increasing Md\textsubscript{30} (\textit{i.e.} increasingly unstable austenite) with increasing IA duration as opposed to Figure \ref{fig:fig-kp-annealing-time-and-shr}a which showed a decreasing $k_p$ with increasing IA duration (\textit{i.e.} increasingly stable austenite). The most likely reason is the inaccurate determination of C. From Figure \ref{fig:fig-ebsd-undeformed} it was established that the austenite fraction at 660 \degree C for short annealing durations was likely much larger than the observed fraction at room temperature. It would therefore be incorrect to use the lever rule with the room temperature phase fractions. While the C content can no longer be determined using the lever rule, the increasing Mn content in the austenite phase with IA duration as seen in Table \ref{tab:austenite compositions} gives confidence that the Md\textsubscript{30} temperature should also be decreasing according to Equation \ref{eq:Md equation}.

\begin{table*}[h!]
	\centering
	\caption{Compositions of austenite phase from various samples measured by TEM-EDS. Standard errors given in parantheses. ND - Not Detected. $\dagger$ C content calculated based on the lever rule and on phase fractions obtained by EBSD. $\ddagger$ Composition of 660-0min sample assumed to be equal to the bulk composition. *Assuming an austenite grain size of 2.9 $\mu$m.  }
	\begin{tabular}{lccccccccccc}
		\toprule
		& Fe    & Mn    & Al    & Si    & C$\dagger$     & Nb    & V     & d$_\gamma$ & SFE   & Md\textsubscript{30} \\
		& \multicolumn{7}{c}{(wt\%)}                            & ($\mu$m)   & (mJ m$^{-2}$)  & (\degree C) \\
		\midrule
	
		660-0 min$\ddagger$ & Bal   & 8.2   & 2.3   & 1.3   & 0.36 & 0.02  & 0.03  & 0.4   & 12.9  & 318 \\
		660-10 min 					& Bal 	& 8.2 (0.1)	& 5.3 (0.2) & 2.4 (0.1)	& 0.69    & 0.03 (0.02) & 0.06 (0.03)   & 2.1	& 46.5	& 164 \\
		660-30 min & Bal   & 10.1 (0.1) & 2.6 (0.1) & 2.0 (0.1) & 0.62 & ND    & 0.04 (0.03) & 2.7   & 26.1  & 185 \\
		660-2 h & Bal   & 11.1 (0.1) & 1.5 (0.1) & 2.2 (0.1) & 0.50 & ND    & 0.03 (0.02) & 2.8   & 13.7  & 231  \\
		\smallskip
		660-24 h & Bal   & 11.9 (0.1) & 2.1 (0.1) & 1.8 (0.1) & 0.48 & ND    & 0.04 (0.02) & 2.9   & 19.1  & 239 \\
		Thermo-Calc & Bal   & 12.5  & 1.6   & 1.6   & 0.48 & trace & trace & -    & 16.1  & 234* & \\
		\bottomrule
	\end{tabular}%
	\label{tab:austenite compositions}%
\end{table*}%

\section{Discussion}

The results in this study have largely confirmed the initial hypothesis to be true, that varying the IA duration was able to vary the strain hardening behaviour of a medium Mn steel. Figure \ref{fig:fig-kp-annealing-time-and-shr} showed that it was possible to reliably predict the $k_p$ value and therefore the $\Theta_{max}$ of the investigated steel with $\Theta_{max}$ showing a linear relationship with the inverse square root of IA duration. Therefore, depending on the application, the investigated steel could be annealed in a CAL for a short duration to produce a stronger steel with a high $\Theta_{max}$ or annealed in a BAF for a longer duration to produce a steel with a lower $\Theta_{max}$ but with more ductility.

The reason behind the change in $\Theta_{max}$ with IA duration can be attributed to the change in austenite stability. Herrera \textit{et al.} \cite{Herrera2011} described that in order to enable a ductile TRIP steel, the austenite should have a sufficiently low stability to enable TRIP but also high enough such that TRIP occurs over a wider strain regime. In this study, the austenite stability was quantified using two different parameters: $k_p$ and Md\textsubscript{30}. The $k_p$ parameter was a more phenomenological approach, comparing the amount of transformed austenite at a given strain. On the other hand, Md\textsubscript{30} attempts to determine the austenite stability from the composition and grain size. While Md\textsubscript{30} is more useful for alloy design, it is still based on empirial data and regression analysis \cite{Angel1954,Nohara1977}. Ultimately, austenite stability depends on many different factors such as composition, grain size, neighbouring constituents and texture \cite{Zhang2013,Xu2017}, making it difficult to determine from first principles. In this study, it is acknowledged that there was some difficulty in measuring the Md\textsubscript{30} temperature accurately and it was the phenomenological parameter $k_p$ which showed a good correlation with the tensile properties of the investigated steel.

As mentioned, austenite stability depends on a variety of microstructural and compositional factors. Considering the microstructural factors in the investigated steel, it could be seen from Figure \ref{fig:fig-ebsd-undeformed} and Table \ref{tab:austenite compositions} that the austenite morphology and grain size did not vary significantly. From Table \ref{tab:phasefracs}, the phase fractions of the 660-2 h, 4 h and 24 h samples were also very similar. As such, these microstructural features were unlikely to have a significant impact on the austenite stability. However, from Table \ref{tab:austenite compositions}, it can be seen that the Mn content in austenite continued to increase from 10 min to 24 h. Even after 24 h of intercritical annealing, the Mn content still did not reach the thermodynamic equilibrium as predicted by Thermo-Calc. This shows that while the steel may be at phase equilibrium, the constituent phases may not be at compositional equilibrium. Mn, which is an austenite stabiliser, also strongly stabilises the austenite phase against strain induced martensitic transformation as shown in Equation \ref{eq:Md equation}. Therefore, in the investigated steel, it is highly likely that the increasing austenite stability with IA duration was largely brought about through continuous Mn partitioning to the austenite phase, \textit{i.e.} chemical stabilisation.

In addition to dislocation glide, the austenite phase in medium Mn steels are known to deform \textit{via} TWIP$+$TRIP or just TRIP, largely depending on the SFE \cite{Lee2014,Field2018,Kwok2022b}. From Table \ref{tab:austenite compositions}, the SFE calculated from the 660-24 h sample was 19.1 mJ m$^{-2}$, however this is expected to be lower (between 15.3-19.1 mJ m$^{-2}$) due to the likely over-measurement of the Al content. Nevertheless, the SFE value of the 660-24 h sample should place the steel in the proposed regime for TWIP$+$TRIP \cite{Lee2014,Kwok2022b}. However, no austenite twins were observed in the postmortem samples under TEM (Figure \ref{fig:fig-tem}). The investigated steel was therefore purely a TRIP-type medium Mn steel, which may explain the good correlation between $k_p$ and $\Theta_{max}$ since there were no other plasticity enhancing mechanisms involved. One reason why twinning did not occur in the investigated steel might be the residual high dislocation density associated with lath-type microstructures \cite{Sun2019a} which is known to suppress twinning \cite{Christian1995}.

From Figure \ref{fig:fig-tem}, twin-like features closely resembling twinned martensite was found in the postmortem 660-10 min sample. However, twinned martensite could also be associated with athermal martensite, \textit{i.e.} formed during cooling \cite{Krauss1971}. Since it was determined that samples intercritically annealed for less than 24 h might contain athermal martensite, it could not be established if the twinned martensite formed during deformation or cooling after IA. However, twinned martensite was still found in the 660-24 h postmortem sample as shown in Figure \ref{fig:fig-tem}g-i. Since athermal martensite was not expected in this sample, this strongly suggests that the observed twinned martensite was the product of deformation induced martensitic transformation. \textcolor{black}{Medium Mn steels have shown to be able to form a variety of different types of martensites such as strain induced martensite \cite{Lee2014} and stress-assisted martensite \cite{Yen2015}. However, the significance and role of twinned martensite in medium Mn steels are not yet clearly understood, although the type of martensite gives certain clues as to how it is formed.} The formation of twinned martensite during deformation is not commonly reported in medium Mn steels \cite{Lee2015b,Kwok2022b} but is more often reported in TRIP steels \cite{Samek2006}, TRIP-assisted steels \cite{Min2016} and Fe-Ni-C steels \cite{Maxwell1974,Zhang2001}. Based on the work by Zhang and Kelly \cite{Zhang2001}, and Kelly and Nutting \cite{Kelly1960} in Fe-Ni-C steels, twinned martensite is formed during a deformation induced martensitic transformation when the C content of the austenite is above 0.4 wt\%, which is consistent with the estimated C content of the 660-24 h sample in Table \ref{tab:austenite compositions}.



\section{Conclusions}

A novel concept of varying the IA duration at a single temperature instead of varying the IA temperature was explored. An 8 wt\% Mn medium Mn steel was hot rolled, quenched and intercritially annealed at 660 \degree C for durations between 10 min and 24 h. This produced a range of austenite stabilities because of continuous Mn enrichment of the austenite phase with increasing IA duration. When tested in uniaxial tension, the $\Theta_{max}$ was found to decrease with increasing IA duration. A good linear fit was found between $\Theta_{max}$ and the austenite stability constant $k_p$ and thus between $\Theta_{max}$ and the IA duration.

The deformation structures were observed with TEM and found that TRIP, rather than TWIP$+$TRIP was the dominant plasticity enhancing mechanism despite the calculated SFE being within the correct range for TWIP$+$TRIP. It was postulated that the residual high dislocation density in the microstructure might have suppressed twinning in the steel. The likely formation of twinned martensite was observed and determined to be the primary deformation induced martensitic product in the steel due to the high C content of the prior austenite. 

\section{Acknowledgements}
TWJK would like to thank A*STAR, Singapore for provision of a studentship. DD acknowledges funding from the EPSRC grant Designing Alloys for Resource Efficiency (DARE), EP/L025213/1.

\bibliographystyle{MMTA_Thomas3}
\bibliography{library}

\end{document}